\documentclass[aps,prb,reprint,showpacs,twocolumn,superscriptaddress]{revtex4-1}
\usepackage[utf8]{inputenc}
\usepackage[american,british]{babel}
\usepackage[T1]{fontenc}
\usepackage[pdftex]{graphicx}  
\usepackage{graphicx, xcolor}
\usepackage{dcolumn}
\usepackage{bm}
\usepackage{amsmath,amsthm,amssymb}
\usepackage{color}
\usepackage{verbatim}
\usepackage{algorithm}
\usepackage[noend]{algpseudocode}

\definecolor{darkGreen}{RGB}{0,110,0}
\definecolor{darkBlue}{RGB}{0,0,130}
\usepackage[colorlinks,bookmarks=false,citecolor=blue,linkcolor=red,urlcolor=blue]{hyperref}

\begin{document}
\title{Negativity Spectrum in the Random Singlet Phase}

\author{Xhek Turkeshi}
\affiliation{SISSA and INFN, via Bonomea 265, 34136 Trieste, Italy}
\affiliation{The Abdus Salam International Centre for Theoretical Physics, strada Costiera 11, 34151 Trieste, Italy}
\author{Paola Ruggiero}
\affiliation{SISSA and INFN, via Bonomea 265, 34136 Trieste, Italy}
\author{Pasquale Calabrese}
\affiliation{SISSA and INFN, via Bonomea 265, 34136 Trieste, Italy}
\affiliation{The Abdus Salam International Centre for Theoretical Physics, strada Costiera 11, 34151 Trieste, Italy}

\date{\today}

\begin{abstract}
Entanglement features of the ground state of disordered quantum matter are often captured by 
an infinite randomness fixed point that, for a variety of models, is  the random singlet phase.
Although a copious number of studies covers bipartite entanglement in pure states, at present, less is known for mixed states and tripartite settings. 
Our goal is to gain insights in this direction by studying the negativity spectrum in the random singlet phase. 
Through the strong disorder renormalization group technique, we derive analytic  formulas for the universal scaling of the 
disorder averaged moments of the partially transposed reduced density matrix. 
Our analytic predictions are checked against a numerical implementation of the strong disorder renormalization group and against exact computations for the XX spin chain 
(a model in which free fermion techniques apply). 
Importantly, our results show that the negativity and  logarithmic negativity are not trivially related after the average over the disorder. 
\end{abstract} 

\maketitle

\section{Introduction}
Entanglement is fundamental in understanding quantum phases of matter~\cite{Amico2007,Calabrese2009R,Laflorencie2015}. Mathematically defined as a measure of non-separability on quantum states, its intrinsic non-local nature renders this quantity theoretically and experimentally challenging to measure \cite{Islam2015}.  
Let us first consider a bipartition of a system into two spatial regions ${A\cup B}$, with Hilbert space ${\mathcal{H}=\mathcal{H}_A\otimes\mathcal{H}_B}$, and
a pure state ${|\Psi\rangle \in \mathcal{H}}$ with  reduced density matrix ${\rho_A=\text{tr}_B |\Psi\rangle\langle \Psi|}$. 
The content of bipartite entanglement can be read from the R\'enyi entanglement entropies
\begin{equation}
\label{eq:1.0}
	S_\alpha(\rho_A) = \frac{1}{1-\alpha}\log\text{tr} \left(\rho_A^\alpha\right),
\end{equation}
and from their von Neumann limit
\begin{equation}
	S(\rho_A) =\lim_{\alpha\to 1} S_\alpha(\rho_A)= -\text{tr} (\rho_A\log \rho_A).
\end{equation}
The knowledge of $S_\alpha$ for positive values of $\alpha$ fixes the entire spectrum of $\rho_A$. 
The latter, usually referred to as \emph{entanglement spectrum}, has been proven of fundamental value in a variety of frameworks, including topological properties of quantum matter~\cite{Li2008,Regnault2009,Fidkowski2010,Lauchli2010,Yao2010,Pollmann2010B,Dubail2011,Qi2012,Poilblanc2012,Cincio2013,Lundgren2016}, 
symmetry-broken phases~\cite{Metlitski2011,Alba2013,Tubman2014,Kolley2013,Frerot2016} and many-body localization~\cite{Yang2015,Geraedts2016,Serbyn2016}. 
For one-dimensional critical systems with an underlying conformal invariance, the distribution of eigenvalues of ${\rho_A}$ obeys a universal scaling law, 
depending only on the central charge~\cite{Lefevre2008,Lauchli2013,Pollmann2010,Nakagawa2017,gs-18,Alba2018A}.
This distribution is of high importance to understand the effectiveness of some tensor network algorithms \cite{Tagliacozzo2008,Pollmann2009,Pirvu2012}.

The situation is more complicated when considering a bipartition of a system in a mixed state. 
Here the R\'enyi entropies~\eqref{eq:1.0} do not distinguish between classical and quantum correlations, and thus, they fail to characterize entanglement. The same issue arises when considering the mutual entanglement between subregions of a multipartite pure state. 
For concreteness, let us consider a tripartition $A_1 \cup A_2 \cup B$ of a pure state 
${|\Psi\rangle \in\mathcal{H} = \mathcal{H}_{A_1}\otimes \mathcal{H}_{A_2}\otimes \mathcal{H}_B}$. 
Tracing out $B$, we obtain the reduced density matrix $\rho_A$ describing the subsystem ${A=A_1\cup A_2}$. 
The quantum correlations between  $A_1$ and $A_2$ are encoded in the partially transposed reduced density matrix~\cite{Peres1996,Horodecki1996,Simon2000,Werner2001,Giedke2001,Zyczkowski1998,Zyczkowski1999} $\rho_A^{T_2}$ and in its negative eigenvalues. 
The definition of $\rho_A^{T_2}$ is ${\langle u v | \rho_A^{T_2} | u' v'\rangle \equiv \langle u v' | \rho_A | u' v \rangle}$, with $\{u\}$ and $\{v\}$ being local bases of 
respectively ${\mathcal{H}_{A_1}}$ and ${\mathcal{H}_{A_2}}$.
From $\rho_A^{T_2}$ one can extract measures of mutual entanglement such as the entanglement negativity ${\cal N}$ and the logarithmic negativity 
${\cal E}$~\cite{Eisert1999,Lee2000,Vidal2002,Plenio2005}
\begin{equation}
\label{eq:1.1}
\mathcal{N}  = \frac{\big|\big| \rho_A^{T_2}\big|\big| -1 }{2}, \quad  \mathcal{E}=\log\big|\big| \rho_A^{T_2}\big|\big|,
\end{equation}
where ${\big|\big| \rho\big|\big|=\text{tr}\sqrt{\rho \rho^\dagger}}$ is the trace norm. 
For any given  state, clearly ${\cal E}=\log (2 {\cal N}+1)$.
The (logarithmic) negativity has been studied in several contexts, ranging from  harmonic chains and 
lattices~\cite{Audenaert2002,Ferraro2008,Cavalcanti2008,Anders2008,Anders2008B,Marcovitch2009,Sherman2016,Nobili2016,ez-16c,mm-18} 
to quantum spin models~\cite{Wichterich2009,Bayat2010,Bayat2010B,Bayat2012,Wichterich2010,Santos2011,Grover2018,Javanmart2018,gbpb-19,cgs-19,sdhs-16,ksr-19}, 
from conformal and integrable field theories~\cite{Calabrese2012B,Calabrese2013B,Calabrese2013C,Alba2013C,Coser2014,Calabrese2015B,Nobili2015,Kulaxizi2014,Fournier2016,Bianchini2016} 
to non-equilibrium situations \cite{Coser2014,Eisler2014B,Alba2018,Hoogeveen2015,Wen2015,Gullans2019,knr-19} and intrinsic and symmetry-protected topological 
orders~\cite{Wen2016A,Wen2016B,Castelnovo2013,Lee2013,Hart2018,Pollmann2012C,Shiozaki2017,Shapourian2017,Shiozaki2017,Shiozaki2018}. 
For fermionic models, it has been shown that the partial time-reversal transpose is a more appropriate object to characterise the 
entanglement in mixed states \cite{ez-15,Shapourian2017,Shiozaki2018,cw-16,hw-16,eez-16,ssr-17,sr-19,sr-19b,ctc-16}.
Finally, also experimental proposals for the measurement of negativity have recently appeared~\cite{Gray2018,Cornfeld2018}.

It is not a surprise that the spectral density of ${\rho_A^{T_2}}$ contains more information about the entanglement between $A_1$ and $A_2$ than the negativities 
in Eq.~\eqref{eq:1.1}. Such spectral density is usually referred to as \emph{negativity spectrum} \cite{Ruggiero2016} and 
is fully characterized through the moments
\begin{equation}
\label{eq:1.2}
	M_\alpha^{T_2} = \text{tr} \left(\rho_A^{T_2}\right)^\alpha.
\end{equation}
In the following we will refer to $M_\alpha^{T_2}$ as {\it negativity moments}. 
Notice that the  trace norm in Eq.~\eqref{eq:1.1} may be obtained as the replica limit\cite{Calabrese2012B,Calabrese2013B}
${\big|\big| \rho_A^{T_2} \big|\big|=\lim_{\alpha\to 1/2}  M_{2\alpha}^{T_2}}$.

The negativity spectrum so far has been investigated only for clean systems \cite{Ruggiero2016,Mbeng2017,Hassan2019A}.
On the other hand, when considering quenched disorder, static and dynamic properties of a system drastically change compared to the clean case. In fact, randomness usually plays a relevant role in the renormalization group sense. Remarkable examples are Anderson and many-body localization~\cite{Anderson1958,Abrahams2010,Nandkishore15,Abanin2019}. Other well studied systems include a class of quantum spin chains where disorder induces a novel quantum critical phase~\cite{Ma1979,Ma1980,Fisher92,Fisher92B,Igloi05,Monthus2018}. This phase is characterized by the formation of spin singlets spreading over arbitrarily large distances, and for this reason, it is dubbed random singlet phase (RSP). 
Its features can be analytically accessed by the strong disorder renormalization group (SDRG) technique~\cite{Igloi05,Monthus2018}. 
Concerning entanglement, it was found that the disorder-averaged entanglement entropy of the RSP follows a universal scaling law~\cite{Refael2004,Laflorencie2005,Dechiara2006,Refael2009}. Similar results have been derived in other disordered fixed points and singlet phases~\cite{Raul2006,Refael2007,Hoyos2007,Lin2007,Bonesteel2007,Binosi2007,Igloi2008,Fidkowski2008,Yu2008,Hoyos2011,Kovaks2009,Kovaks2012,Getelina2016,Laguna2016,Burrell2007,Igloi2012,Bardarson2012,asr-18,pcp-19}. 
Furthermore, the disorder-averaged entanglement spectrum~\cite{Pouranyari2013,tmd-18} and its moments~\cite{Fagotti2011} have been studied, as well as the low-lying excitations~\cite{Ramirez2014}.

In the same fashion, the disorder-averaged logarithmic negativity displays a universal scaling law~\cite{Ruggiero2016B}, but 
the negativity spectrum in disordered systems has not been studied yet. 
This work provides a first analysis on the subject, focusing on the random singlet phase. In the spirit of Ref.~\onlinecite{Fagotti2011} we use renewal equations to find analytic formulas for the negativity moments. In particular, we work out analytic result for the the case of adjacent intervals which we test 
against a numerical implementation of the SDRG  and against  \emph{ab-initio} simulations for the random XX spin-chain. 
Among the other results, we find that the logarithm and the average disorder do not commute, in the sense that 
${\overline {\log M^{T_2}_{2\alpha}}\neq {\log \overline{M^{T_2}_{2\alpha}}}}$ even in the limit ${\alpha\to1/2}$.
A maybe surprising consequence is that the negativity and its logarithmic analogue are not trivially related after disorder average in the RSP.
This is in contrast with the case of the entanglement entropy, where the logarithm and the disorder average commute in the replica limit ${\alpha \to 1}$\cite{Fagotti2011}.

The remaining of the paper is organized as follows. In Sec.~\ref{sec:RSP} we review the strong disorder renormalization group and the random singlet phase for the systems of interest. In Sec.~\ref{sec:ENS}, we explain how the negativity spectrum can be characterized by the negativity moments. We then introduce the renewal equation for the negativity generating function and work out the negativity moments for adjacent intervals. The analytic solutions are benchmarked numerically in Sec.~\ref{sec:NUM}. In the last section we discuss the obtained results and possible outlooks. 
In an appendix we report the results for the fermionic negativity moments of the same disordered model.

\section{Random singlet phase}
\label{sec:RSP}
The random singlet phase is the simplest infinite-randomness fixed point~\cite{Igloi05,Monthus2018,Refael2009}. It describes, for example, the low-energy properties of the spin-1/2 disordered Heisenberg and XX chains, which are particular instances (respectively ${\Delta=1}$ and ${\Delta=0}$) of the random XXZ chain, with Hamiltonian
\begin{equation}
\label{eq:2.0}
	H = \sum_{i=1}^{L-1} J_i(\sigma_i^x \sigma_{i+1}^x+ \sigma_i^y \sigma_{i+1}^y + \Delta \sigma_i^z \sigma_{i+1}^z).
\end{equation}
Here ${\sigma_i^\alpha}$ (${\alpha=x,y,z}$) denotes the Pauli matrices at site $i$, and the $\{J_i\}$ are positive uncorrelated quenched random couplings drawn by a probability distribution $P(J)$. It has been shown that the low-energy/long-distance properties are disorder independent, i.e., they are the same for essentially any choice of $P(J)$~\cite{Fisher92}. In the numerical section of this paper, we exploit this freedom by restricting to the uniform distribution ${P(J)=1}$, with ${J\in [0,1]}$.

The random singlet phase emerges from the SDRG applied to Eq.~\eqref{eq:2.0}. In this section we briefly review and discuss some useful properties. 

\subsection{Universality of the phase}
For disordered spin systems, the usual space-block decimation~\cite{Cardybook} fails due to the inhomogeneity of the hamiltonian \eqref{eq:2.0} within a single disorder realization. The rationale is instead to decimate through an energetic principle, where, at each renormalization step, the sites connected by the strongest coupling are projected onto their local ground state, i.e., the singlet state. These are effectively decoupled by the rest of the system, while the edge sites are connected by a renormalized coupling. 

For concreteness, let us focus on the random Heisenberg chain, although a similar procedure holds also for the XX chain. 
We denote the strongest bond by ${\Omega=J_j=\max_i J_i}$ (for some $j$) and rewrite the Hamiltonian as ${H=H_\Omega+H_\textup{edge} + H_\textup{rest}}$, where
\begin{align}
\label{eq:2.1a}
	& H_\Omega  =\Omega \vec{\sigma}_j\cdot\vec{\sigma}_{j+1},\\
\label{eq:2.1b}
	& H_\textup{edge}  = J_\textup{L} \vec{\sigma}_{j-1}\cdot\vec{\sigma}_{j}+J_\textup{R} \vec{\sigma}_{j+1}\cdot\vec{\sigma}_{j+2},\\
\label{eq:2.c}
	& H_\textup{rest}  = \sum_{i\neq j-1,j,j+1} J_{i}\vec{\sigma}_i\cdot\vec{\sigma}_{i+1}.
\end{align}
The first line gives the Hamiltonian of the strongest bond  connecting the sites ${(j,j+1)}$. 
The Hamiltonian ${H_\textup{edge}}$ represents the interaction of these sites with the neighboring spins, while the last equation is the Hamiltonian of all the other degrees of freedom. 
For ${\Omega}$ positive, the ground state of $H_\Omega$ is the singlet state
\begin{equation}
	|s_j\rangle \equiv \frac{|\uparrow_j\downarrow_{j+1}\rangle-|\downarrow_j\uparrow_{j+1}\rangle}{\sqrt{2}}.
\end{equation}
These two sites forming a singlet can be now decoupled, while the edge spins ${(j-1,j+2)}$ interacts via an effective Hamiltonian $H_\textup{eff}$, 
obtained through second order perturbation theory in $1/\Omega$. Apart from an unimportant additive constant, this reads
\begin{equation}
	\label{eq:2.2}
	H_\textup{eff} = \tilde{J}_{j-1} \vec{S}_{j-1}\cdot\vec{S}_{j+2},\qquad \tilde{J}_{j-1} = \frac{J_\textup{L} J_\textup{R}}{2\Omega}.
\end{equation}
After this renormalization step, the Hamiltonian ${H'=H_\textup{eff}+H_\textup{rest}}$ is of the same form as the initial $H$, and the procedure can be iterated. The single step is called Ma-Dasgupta rule~\cite{Ma1979,Ma1980} and can be summarized as 
\begin{equation}
\label{eq:2.3}
	\left(\dots,J_\textup{L},\Omega,J_\textup{R},\dots\right)_L\to \left(\dots,\frac{J_\textup{L}J_\textup{R}}{2\Omega},\dots\right)_{L-2}.
\end{equation}
In the last equation, we specified that the chain length reduced from $L$ to $L-2$ in one renormalization step. 
We stress that the SDRG results are valid in the ${L\to\infty}$ limit, and finite size corrections are present when numerically implementing Eq.~\eqref{eq:2.3} (see Sec.~\ref{subsec:nsdrg}). 

\begin{figure}[t]
	\includegraphics[width=\columnwidth]{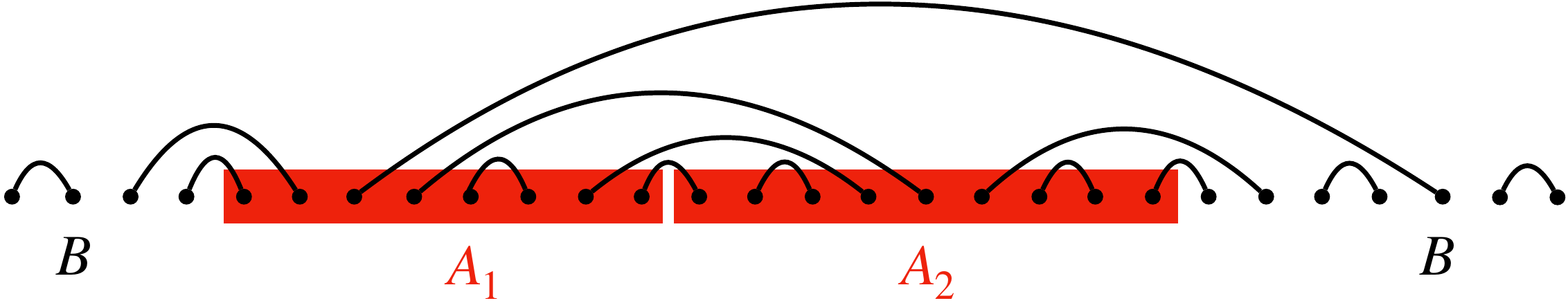}
\caption{\label{Fig.1}
Cartoon of the random singlet phase in a tripartite setting ${A_1 \cup A_2 \cup B}$. The special case of two adjacent intervals (in red) embedded in a larger system is depicted.}
\end{figure}

Successive applications of the Ma-Dasgupta rule lead asymptotically to a product state of singlets at arbitrarily large distances, the so-called random singlet phase (RSP), depicted in Fig.~\ref{Fig.1}. Here, singlets between more distant sites are generated at later SDRG steps.  

In order to understand how universality emerges in the RSP phase, it is convenient to introduce the  variables
\begin{equation}
	\label{eq:2.4}
	\beta^{m}_i = \log \frac{\Omega^{(m)}}{J_i^{(m)}},\quad {\Gamma^{(m)} = -\log \Omega^{(m)}}.
\end{equation}
Here ${\Omega^{(m)}}$ and ${J_i^{(m)}}$ are respectively the strongest bond and the couplings at site $i$ at renormalization step $m$. Intuitively, ${\Gamma}$ set the energy scale of the strongest coupling at a successive step, while ${\beta}$ is a measure of the broadness of the coupling distribution around it.

The Ma-Dasgupta rule \eqref{eq:2.3} rewritten in terms of the $\beta$ variables is ${\tilde{\beta} =\beta_\textup{L} + \beta_\textup{R}-\log{2}}$. 
It induces a flow for the  probability distribution of the couplings ${P(\beta,\Gamma)}$
\begin{align}
\label{eq:2.6}
	\frac{d}{d\Gamma}&P(\beta,\Gamma) = \frac{\partial}{\partial\beta}P(\beta,\Gamma) + P(0,\Gamma)\int_0^\infty d\beta_2 \int_0^\infty d\beta_1\nonumber\\
	&\times \delta(\beta_1+\beta_2-\log{2}-\beta)P(\beta_1,\Gamma)P(\beta_2,\Gamma).
\end{align}
Iterating the renormalization procedure, $\beta$ grows indefinitely and it is safe to drop out the factor $\log2$ in the above equation. 
Within this assumption, Eq.~\eqref{eq:2.6} can be solved analytically~\cite{Fisher92}, leading to
\begin{equation}
\label{eq:2.7}
	P_\star(\beta,\Gamma) = \frac{e^{-\beta/\Gamma}}{\Gamma}.
\end{equation}
This function is a \emph{universal attractor}, irrespective of the distribution of the couplings~\cite{Fisher92,Fisher92B,Refael2009}. Moreover, variables distributed according to Eq.~\eqref{eq:2.7} are closely packed around ${J_\star\simeq 0^+}$, and this \emph{a posteriori} justifies the perturbative treatment.

We close by recalling that similar results hold for the XX chain, where the Ma-Dasgupta rule reads \cite{Igloi05}
\begin{equation}
	\left(\dots,J_\textup{L},\Omega,J_\textup{R},\dots\right)_L\to \left(\dots,\frac{J_\textup{L}J_\textup{R}}{\Omega},\dots\right)_{L-2}.
\end{equation}
It is evident that the random XX and Heisenberg chains belong to the same universality class since they share the same fixed point distribution ${P_\star(\beta,\Gamma)}$.

\subsection{Structure of the reduced density matrix and its partial transpose}
\label{subsec:structureRDM}

The RSP emerges naturally as an infinite disorder critical point, and it is characterized by singlets spreading among arbitrary far regions of the system. 
Below we introduce the elementary building blocks of the associated density matrix and its partially transpose. 
They are (i) the density matrix of a singlet, $\rho_\textup{2s}$, (ii) its reduced density matrix for one of the spins, ${\rho_\textup{s}}$, (iii) the partial transpose of $\rho_\textup{2s}$ with respect to one of the sites, ${\rho_\textup{2s}^{T_2}}$.
In the basis  ${\left|\uparrow\uparrow\right\rangle}$, ${\left|\uparrow\downarrow
\right\rangle}$, ${\left|\downarrow\uparrow\right\rangle}$, and ${\left|\downarrow
\downarrow\right\rangle}$, the above objects read
\begin{align} 
\label{eq:2.1.1}
\rho_\textup{2s} =
&\frac{1}{2}\begin{pmatrix}
0 & 0   &  0 & 0\\
0 & 1   & -1 & 0\\
0 & -1  &  1 & 0\\
0 & 0   &  0 & 0
\end{pmatrix},\quad \rho_\textup{s} =
\frac{1}{2}\begin{pmatrix}
1  & 0 \\
0 &  1  
\end{pmatrix},\\ \label{eq:2.1.2}&\qquad \rho^{T_2}_\textup{2s}=\frac{1}{2}
\begin{pmatrix}
 0 & 0  & 0 & -1 \\
 0 & 1  & 0 & 0 \\
 0 & 0  & 1 & 0 \\
-1 & 0  & 0 & 0 
\end{pmatrix}.
\end{align}
 For concreteness, we consider the partition of the system ${B\cup A_1\cup A_2}$ (pictorially represented in Fig.~\ref{Fig.1}), with ${A=A_1\cup A_2}$. 
 We denote with ${n_{X:Y}}$ the number of singlets shared between $X$ and $Y$. 
 This is symmetric ${n_{X:Y}=n_{Y:X}}$ and additive ${n_{X:Y\cup Z} = n_{X:Y}+n_{X:Z}}$. The density matrix of the RSP takes the form
\begin{equation}
	\rho_\textup{RSP}= \bigotimes_{m=1}^{n_{A:A}}\rho_\textup{2s}\bigotimes_{n=1}^{n_{A:B}}\rho_\textup{2s}\bigotimes_{l=1}^{n_{B:B}}\rho_\textup{2s}.
\end{equation}
Tracing out $B$ we obtain
\begin{equation}
	\rho_A = \bigotimes_{m=1}^{n_{A:A}}\rho_\textup{2s}\bigotimes_{n=1}^{n_{A:B}}\rho_\textup{s},
\end{equation}
whose partial transpose with respect to $A_2$ gives
\begin{equation}
\label{eq:2.A}
	\rho_A^{T_2}=\bigotimes_{q=1}^{n_{A_1:A_1}}\rho_\textup{2s}\bigotimes_{r=1}^{n_{A_2:A_2}}\rho_\textup{2s}\bigotimes_{p=1}^{n_{A_1:A_2}}\rho^{T_2}_\textup{2s}\bigotimes_{n=1}^{n_{A:B}}\rho_\textup{s}.
\end{equation}
Here we have used the fact that ${\rho_\textup{s}^{T_2}=\rho_\textup{s}}$ for a single site, and that ${\rho_\textup{2s}^{T_2}=\rho_\textup{2s}}$ when both the ends of a bond are in the same subsystem $A_i$ $(i=1,2)$. The spectrum of Eq.~\eqref{eq:2.A} is denoted as negativity spectrum and is the main object of study in this paper. 

\subsection{Scaling of the in-out bond}
\label{subsec:scalinginout}

The density matrix of a single random configuration and all the  quantities that can be derived from it are fully characterized by the number of in-out bonds $n_{X:Y}$.
Consequently, the scaling of these quantities is crucial in the study of the spectrum of the reduced density matrix of the RSP and of its partial transpose. 
The knowledge of all the $n_{X:Y}$ can be extracted through the solution of a simple set of linear equations, 
relying on the additivity property of  $n_{X:Y}$. 
Hereafter we denote by $\ell_X$ the length of an interval $X$, and ${{X}_c}$ its complement.
Consider a $2k$-multipartite system ${\bigcup_{X\in \mathcal{G}_0}}X$ with ${\mathcal{G}_0 =\{A_1,B_1,\dots,A_k,B_k\}}$. We define ${\mathcal{G}}$ as the set of all possible compact subintervals of the chain. For each ${X\in \mathcal{G}}$ one can decompose the number of singlets $n_{X: X_c}$ as
\begin{equation}
	n_{X:X_c} = \sum_{Y,Z\in \mathcal{G}} n_{X\cap Y: X_c\cap Z}.
\end{equation}
After taking the disorder average, we have a set of linear equations, whose solution gives $\overline{n}_{X:W}$ for any ${X,W\in \mathcal{G}_0}$
\begin{equation}
\label{eq:1.1.B}
	\overline{n}_{X:X_c} = \sum_{Y,Z\in \mathcal{G}} \overline{n}_{X\cap Y: X_c\cap Z}.
\end{equation}
The left hand side has been previously computed within the RSP~\cite{Refael2004}
\begin{equation}
\label{eq:moore}
	{\overline{n}_{X:X_c}=\frac{b_{X:X_c}}{6} \log \ell_X+k},
\end{equation}
where $b_{X:X_c}$ is the number of edges shared by $X$ and $X_c$, and  $k$ is a non-universal constant of order ${\mathcal{O}(1)}$ in the subsystem size $\ell_X$. 
For the leading logarithmic term, we then have
\begin{equation}
\label{eq:1.B}
	\frac{b_{X:X_c}}{6}\log\ell_X = \sum_{Y,Z\in \mathcal{G}} \overline{n}_{X\cap Y: X_c\cap Z}.
\end{equation}
This set of equations can be straightforwardly solved for the variables $\overline n$, and a unique solution can be extracted for any partition of the system. 
See Ref. \onlinecite{Ruggiero2016B} for several explicit examples. 

\section{Negativity spectrum}
\label{sec:ENS}
\subsection{Logarithmic negativity and negativity moments}
The central object of this paper is the spectral density of the operator ${\rho_A^{T_2}}$ in Eq.~\eqref{eq:2.A}
\begin{equation}
\label{eq:ehi}
	\mathbb{P}(\lambda) = \sum_{i} \delta(\lambda-\lambda_i),
\end{equation}
where the sum is over the eigenvalues of ${\rho_A^{T_2}}$. From the knowledge of ${\mathbb{P}(\lambda)}$, we can infer the negativity moments~\eqref{eq:1.2}
\begin{equation}
	\text{tr}(\rho_A^{T_2})^\alpha =\sum_i \lambda_i^\alpha = \int d\lambda\, \mathbb{P}(\lambda)\lambda^\alpha.
\end{equation}
The converse is also true, in that the knowledge of all the negativity moments gives access to the function ${\lambda \mathbb{P}(\lambda)}$ through an inverse Stieltjes 
transform~\cite{Ruggiero2016,Hassan2019A}. For this reason, with a slight but standard abuse of language, we will refer also to the whole set of moments 
as the negativity spectrum.

Most of the derivations presented in this section are valid for a very general tripartition  $A_1\cup A_2\cup B$ of  an infinite chain (with some caveat which will be 
clearer in the course of the calculation).
At the very end of the section, for concreteness, we will specialize to the usual partition depicted in Fig. \ref{Fig.1} with two adjacent blocks. 

Within a single disorder realization, the negativity moments depends only on ${n_{A:B}}$ and ${n_{A_1:A_2}}$. 
The partial transpose ${\rho_A^{T_2}}$ in Eq.~\eqref{eq:2.A} is straightforwardly diagonalized and the 
eigenvalues are
\begin{equation}
\label{eq:2.A.1}
	\lambda_{\pm} = \pm  2^{-n_{A: B} - n_{A_1 : A_2}},
\end{equation}
with degeneracies 
\begin{align}
\label{eq:2.A.2}
	 d_-&= 2^{n_{A:B}+n_{A_1:A_2 } -1}(2^{n_{A_1 : A_2 }} -1), \\ 
	 d_+&= 2^{n_{A: B} + 2 n_{A_1 : A_2}} - d_-. 
\end{align}
Consequently, the negativity moments for this given disorder realization are
\begin{equation} 
\label{eq:2.A.3}
M^{T_2}_\alpha 
= 2^{(n_{A:B} + n_{A_1: A_2}) (1-\alpha)}
\begin{cases}
 2^{n_{A_1: A_2}}  &   \alpha \,\text{even}\\
1                  &   \alpha \, \text{odd}. 
\end{cases}
\end{equation}
Notice that the moments $M^{T_2}_\alpha$ depends on both $n_{A: B}$ and  $n_{A_1 : A_2}$. 
Hence, as well known, they are not direct measures of the mutual entanglement between $A_1$ and $A_2$.
However, the dependence on $n_{A: B}$ cancels in the limit $2\alpha\to1$, as a consequence of the fact
that the negativity is a good entanglement measure also in the RSP\cite{Ruggiero2016B}. 
Nevertheless, in the same spirit of the entanglement spectrum compared to the entanglement entropy\cite{Li2008}, 
the moments~\eqref{eq:2.A.3} encode more information about the mutual entanglement than the (logarithmic) negativity itself, as we shall see. 

Till now we have been discussing what happens for a single disorder realization, but the physical relevant quantities are the averages over the quench disorder. 
From the knowledge of the moments, we can define two different averaged quantities, each one providing useful information about the entanglement.
Indeed, since the average of the logarithm and the logarithm of the average are not at all equivalent, we can define
\begin{align}
\label{eq:2.A.4} 
\hat{\mathcal{E}}_\alpha & = \overline{\log M_\alpha^{T_2}},\\
\label{eq:2.A.4B} {\mathcal{E}}_\alpha &= \log \overline{M_\alpha^{T_2}}. 
\end{align}
These two quantities are expected to behave very differently, as it happens for the analogous averages 
for the entanglement spectrum \cite{Fagotti2011} (i.e. $\overline{\log {\rm tr} \rho_A^\alpha}$ and $\log \overline{{\rm tr} \rho_A^\alpha}$). 
Anyhow, we are going to show that $\hat{\mathcal{E}}_\alpha$ and  ${\mathcal{E}}_\alpha$ are related through a linear transformation at the leading order in $\ell$.

\subsection{Moments $\hat{\mathcal{E}}_{\alpha}$ and logarithmic negativity}
We start by considering  the average of the logarithm of the moments ${\hat{\mathcal{E}}_{\alpha}}$ ($\alpha >0$) in \eqref{eq:2.A.4}. 
This is the easiest quantity to calculate because it depends linearly on $ \overline{n}_{A:B}$ and $\overline{n}_{A_1: A_2}$. 
Hence, straightforwardly from Eq. \eqref{eq:2.A.3}, we get
\begin{equation}
\label{eq:paola1}
\hat{\mathcal{E}}_{\alpha} =
\log 2 \begin{cases}
 (1-\alpha) \overline{n}_{A:B}   + (2-\alpha) \overline{n}_{A_1: A_2} &   \alpha \,\text{even,}\\
  (1-\alpha)(\overline{n}_{A:B}  + \overline{n}_{A_1: A_2})  &   \alpha \,\text{odd}.
 \end{cases}
\end{equation}
We observe that, because of the linear structure, $\hat{\mathcal{E}}_{\alpha} $ depends only on the averages  $\bar{n}_{X:Y}$  and not on the full distribution of the singlets
shared between the partitions.
We recall that one of the main reasons why we are interested in $\hat{\mathcal{E}}_{\alpha} $ is that they are the replica quantities to access the 
average  logarithmic negativity~\cite{Ruggiero2016B}
\begin{equation}
	\label{eq:2.A.5}
	\mathcal{E} = \lim_{\alpha\to 1/2} \hat{\mathcal{E}}_{2\alpha} = \overline{n}_{A_1:A_2} \log 2.
\end{equation}
We stress that \eqref{eq:paola1} are valid for arbitrary tripartition of the chain and not only for adjacent intervals.
Notice that since the moments $\hat{\mathcal{E}}_{\alpha}$ depend only on the averages $\bar{n}_{X:Y}$, they do not encode more information than
the entanglement negativity and entropy. 

\subsection{Moments ${\mathcal{E}}_{\alpha}$ and renewal equation for the negativity spectrum}
\label{subsec:renewalEQ}

The logarithm of the average of the moments in Eq.~\eqref{eq:2.A.4B} is  the quantity more directly related to the true negativity spectrum (i.e. the distribution 
of eigenvalues of the partial transpose). 
Its calculation is, however, much more cumbersome compared to $\hat{\mathcal{E}}_{\alpha}$ because of the non-linear dependence on $n_{X:Y}$: 
it requires the knowledge of the entire distribution of singlets and not only of the average.
We focus on the tripartition ${A=A_1\cup A_2}$ and ${B=A_c}$.
Denoting as ${P(n_{A:B},n_{A_1:A_2})}$ the joint probability distribution of $n_{A:B}$ and $n_{A_1:A_2}$, we  introduce the generating function
for the probability distribution of in-out bonds  \cite{Fagotti2011,Vasseur2015}. 
\begin{multline}
	g(t,s) \equiv \log \langle e^{t n_{A:B} +sn_{A_1:A_2}}\rangle =\\
	= \log \sum_{n_{A:B},n_{A_1:A_2}} e^{n_{A:B}t + n_{A_1:A_2}s} P(n_{A:B},n_{A_1:A_2}) 
	\label{eq:2.C}.
\end{multline}
The knowledge of $g(t,s)$ is equivalent  to the that of  the negativity spectrum, in the sense that it 
univocally determines the negativity moments.

The asymptotic behavior of generating function $g(t,s)$ (in a RG sense that will be clearer later on) may be accessed 
following the phenomenological approach introduced in Ref.~\onlinecite{Fagotti2011} for the entanglement spectrum.
The starting observation is that, within SDRG, the singlets form at a constant rate with respect to the RG time $\mu$. 
This rate is responsible for the logarithmic scaling of ${\overline{n}_{X:X_c}}$ for a single interval $X$. 
The probability distribution of waiting times for a decimation to occur across a bond since the last decimation is \cite{Refael2004}
\begin{equation}
	f(\mu) = \frac{1}{\sqrt{5}} \left(e^{-\frac{3-\sqrt{5}}{2}\mu}-e^{-\frac{3+\sqrt{5}}{2}\mu}\right).
\end{equation}
This expression is true only for asymptotically large $\mu$ because non-universal terms related to the initial distribution of disorder 
have been neglected in its derivation \cite{Refael2004}.
For the following, it is useful to explicitly introduce $\tilde{f}(x)$ as the Laplace transform of $f(\mu)$
\begin{equation}
\tilde{f}(x) = \frac{1}{\sqrt{5}}\left(\frac{1}{x+\frac{3-\sqrt{5}}{2}}-\frac{1}{x+\frac{3+\sqrt{5}}{2}}\right).
\label{tildef}
\end{equation}

At this point, in order to compute $g(t,s)$ one would need to know and quantify all the possible processes between two RG times. 
The renormalization flow generate several of these processes, but the most probable one is clearly the formation of isolated singlets \cite{Fagotti2011}. 
Thus, in a first approximation, expected to be correct in the limit of large $\mu$, we can write a renewal equation for the generating function~\eqref{eq:2.C}, 
considering only formations of in-out isolated singlets
\begin{multline}
	\langle e^{nt+sm}\rangle_\mu = \int_\mu^\infty d\xi f(\xi) + p e^t \int_0^\mu d\xi f(\xi) \langle e^{nt+sm}\rangle_{\mu-\xi} 
	\\ +qe^s \int_0^\mu d\xi f(\xi) \langle e^{nt+sm}\rangle_{\mu-\xi}.
	\label{renew}
\end{multline}
Here, for notational convenience, we express the disorder average at RG time $\mu$ with ${\langle \cdot \rangle_\mu}$ while $n$ and $m$ are just shorthands 
for $n_{A:B}$ and $n_{A_1:A_2}$ respectively. 
The constants $p$ and $q=1-p$ are, respectively, the asymptotic probability of increasing ${n}$ and ${m}$ by one unit. 
In a general setting $p$ and $q$ can depend on the RG time $\mu$ and can have activation times depending on the tripartition, 
here we are only interested in the limit of large $\mu$ and hence neglect these corrections that can be important when comparing with numerics. 
The {\it fundamental assumption} here is that $p$ and $q$ have a non-zero limit as ${\mu\to\infty}$.
The renewal equation \eqref{renew} represents an educated conjecture generalizing the one for $n_{A:B}$ in Ref. \onlinecite{Fagotti2011}
to two kinds of singlets ($n_{A:B}$ and $n_{A_1:A_2}$) with probability $p$ and $q$. 
The correctness of all our (reasonable) assumptions can be tested only a posteriori with numerical simulations.

The renewal equation \eqref{renew}  can be solved through Laplace transform. 
Indeed, after some simple algebra we get
\begin{equation}
	g_\mu (t,s) = \log \left[\mathcal{L}^{-1}\left(\frac{1}{x}\frac{1-\tilde{f}(x)}{1- (p e^t + q e^s)\tilde{f}(x)}\right)(\mu)\right].
	\label{solution}
\end{equation}
The inverse transform $\mathcal{L}^{-1}$ can be computed analytically and gives, at large $\mu$ 
\begin{multline}
	e^{g_\mu(t,s)} \stackrel{\mu\gg 1 }{\simeq} \left(\frac12 + \frac3{2\sqrt{5+4(p e^t+q e^s)}}\right)\times \\
	\label{eq:2.D.1}
	\qquad\qquad \exp{\displaystyle\left( \frac{\sqrt{5+4(pe^t+qe^s)}-3}{2}\mu\right)}.
\end{multline}
From the definition~\eqref{eq:2.C}, we have 
\begin{align}
\label{eq:2.D.2a}
	\overline{n}_{A_1:A_2} = \partial_s g(0,0) &\stackrel{\mu\gg1}{=} \frac{q}{9}(3\mu -1),\\
	\label{eq:2.D.2b}
	\overline{n}_{A:B} = \partial_t g(0,0) &\stackrel{\mu\gg1}{=} \frac{p}{9}(3\mu -1),
\end{align}
and in particular 
\begin{equation}
\label{eq:2.X.1}
	\frac{\overline{n}_{A_1:A_2}}{\overline{n}_{A:B}} = \frac{q}{p}=\frac{q}{1-q}.
\end{equation}
The last three equations must be used to extract $p$ and ${q=1-p}$ in a self-consistent way. 
Indeed, the average number of singlets between complementary sets is univocally fixed by the set of equations~\eqref{eq:1.B}. 
Thus, for a chosen partitioning $\mathcal{G}_0$, one first solves~\eqref{eq:1.B}, then uses the solutions to determine the probabilities  
$p$ and $q$ via Eq.~\eqref{eq:2.X.1}, and finally plug them in Eq. \eqref{solution} determining the asymptotics of $g(t,s)$ for large $\mu$.
Notice that in Eqs.~\eqref{eq:2.D.2a} and~\eqref{eq:2.D.2b} we kept the $O(1)$ term in $\mu$ to show that Eq.~\eqref{eq:2.X.1} is valid also 
at the first subleading order.

At this point, Eq.~\eqref{eq:2.A.3} allows to write the desired averaged negativity moments as function of $g(t,s)$ as 
\begin{equation} 
	\mathcal{E}_\alpha = g(t_\alpha,s_\alpha),
\end{equation}
with
\begin{eqnarray}
t_{\alpha} &\equiv &(1-\alpha)\log 2 ,\\
s_{\alpha}& \equiv&
\begin{cases}
(2-\alpha) \log 2  &   \alpha \,\text{even},\\
(1-\alpha) \log 2  &   \alpha \, \text{odd}. 
\end{cases}
\end{eqnarray}

The leading term in $\mu$ (and hence in $\overline{n}$ or equivalently in $\ell$) comes from the exponential term in Eq.~\eqref{eq:2.D.1}.
We have two different results for $\alpha$ even and odd that we denote respectively as $\mathcal{E}^e_\alpha$ and $\mathcal{E}^o_\alpha$.
By simple algebra we obtain
\begin{equation}
\mathcal{E}^e_\alpha=  \frac{\sqrt{5+(1+q)2^{3-\alpha}} - 3}{2} \mu+\dots, 
\label{Ee}
\end{equation}
and 
\begin{equation}
\mathcal{E}^o_\alpha=  \frac{\sqrt{5+2^{3-\alpha}} - 3}{2} \mu+\dots, 
\label{Eo}
\end{equation}
where the dots stands for subleading non-universal terms in $\mu$.
Notice that in Eqs.~\eqref{Ee} and~\eqref{Eo} all the dependence on the partition is encoded in the constant $q$ and in $\mu$.
However, since $\mu$ is proportional to the logarithm of the length involved in the problem, the universal prefactor of this logarithm 
depends on the partition only through $q$. Hence, the odd moments have the same scaling factor for any tripartition of the chain with ${q\neq0,1}$. 
 
Eqs.~\eqref{Ee} and~\eqref{Eo} are the main analytic results of this manuscript and we recall that {\it they are valid for any tripartition of the infinite chain} 
as long as ${p,q\neq0}$.
They contain a lot of physical insights that we are going to discuss now. 
First of all, they depend on the entire distribution of shared singlets and not only on they averaged values, showing indeed that the negativity moments provide more 
information than the logarithmic negativity and the entanglement entropy. 
A trivial consistency check is that ${{\cal E}^o_1=0}$, as it should.
An important consequence of Eq.~\eqref{Ee} is that the replica limit ${{\cal E}^e_1={\lim_{\alpha\to 1}\mathcal{E}_{2\alpha}}}$ does not converge to the logarithmic negativity Eq.~\eqref{eq:2.A.5} (which is the limit of $\hat{\mathcal{E}}_{2 \alpha}$) for any $q>0$.
This means that \emph{the average negativity is not related trivially to the average logarithmic negativity} as instead happens for a clean system, 
i.e. ${\overline{\cal E}\neq \log (2 \overline{\cal N}-1)}$ as average over disorder. 
Not only, we also have that ${{{\cal E}^e_1> \hat{\cal E}^e_1}}$ for all $q>0$, as expected since the logarithm is a concave function. 
It is also true that ${{{\cal E}_\alpha> \hat{\cal E}_\alpha}}$ for any $\alpha$.

Since both ${\cal E}_\alpha$ and $\hat{\cal E}_\alpha$ are proportional to $\mu$, it is instructive also to write a relation between the two at fixed $\alpha$.
Simply combining Eqs.~\eqref{Ee} and \eqref{Eo} with \eqref{eq:paola1}, we obtain
\begin{equation}
\label{eq:res1}
	\mathcal{E}_\alpha =  A_\alpha \hat{\mathcal{E}}_\alpha + K^{(\alpha)},
\end{equation}
were ${K^{(\alpha)}}$ are a non-universal constants  and function $A_\alpha$ takes two different values for even and odd $\alpha$: 
\begin{eqnarray}
\label{eq:res2}
	A_\alpha^e &=& 3\frac{\sqrt{5+(1+q)2^{3-\alpha}} - 3}{2 (1-\alpha + q)\log 2}, \label{Ae} \\
	A_\alpha^o &=& 3\frac{\sqrt{5+2^{3-\alpha}} - 3}{2 (1-\alpha)\log 2}. \label{Ao}
\end{eqnarray}

\subsection{Application to adjacent intervals}
\label{subsec:adja}

In this subsection we specialize the results of the previous one to the case of adjacent intervals of length $\ell_1$ and $\ell_2$ as in Fig.~\ref{Fig.1}.
In this case the set of equations~\eqref{eq:1.B} admits the following solution at the leading order in the lengths \cite{Ruggiero2016B}
\begin{align}
\label{eq:2.C.2}
	& \overline{n}_{A_1:A_2} = \frac{1}{6}\log\left(\frac{\ell_{A_1}\ell_{A_2}}{\ell_{A_1}+\ell_{A_2}}\right),\\
	\label{eq:2.C.3}
	& \overline{n}_{A:B} = \frac{1}{3}\log{( \ell_{A_1}+\ell_{A_2})}.
\end{align}
The ratio \eqref{eq:2.X.1} seems a complicated function of $\ell_1$ and $\ell_2$.
However, we are interested in the regime of ${\ell_1\propto \ell_2\gg 1}$ when  $ {\overline{n}_{A_1:A_2}/\overline{n}_{A:B}=1/2+\dots}$, 
where the dots stand for subleading logarithmic corrections to the scaling (which may be important in the analysis of the numerical data). 
Hence, in the regime ${\ell_{1,2}\gg1}$, from Eq.~\eqref{eq:2.X.1} we get 
\begin{equation}
q=\frac13, \quad  {\rm and} \quad p=\frac23.
\label{pq}
\end{equation} 

Summarizing, plugging Eq.~\eqref{pq} in Eqs.~\eqref{eq:paola1}, \eqref{Ee} and~\eqref{Eo},
the final results for $\hat{\mathcal{E}}_\alpha $ and ${\mathcal{E}}_\alpha$ for adjacent intervals are
\begin{eqnarray}
	\hat{\mathcal{E}}_\alpha^e &
		=&(4-3\alpha) \frac{\log2}{6}\log{\ell} + \dots,  \label{eq:Z1}\\
\hat{\mathcal{E}}_\alpha^o &=&(1-\alpha) \frac{\log2}{2}\log{\ell} + \dots,\\
{\mathcal{E}}_\alpha^e &=&3\frac{\sqrt{5+ 2^{5-\alpha}/3}-3}4 \log\ell  + \dots,\\
{\mathcal{E}}_\alpha^o &=& 3\frac{\sqrt{5+2^{3-\alpha}} - 3}{4} \log\ell  + \dots,
\label{eq:Z4}
\end{eqnarray}
where we posed ${\ell=\ell_1= x \ell_2}$ (with $x$ finite) and the dots stand (again) for non-universal additive constants (with a partial and universal dependence on $x$).

The relation between $\hat{\mathcal{E}}_\alpha $ and ${\mathcal{E}}_\alpha$ is always given by Eq.~\eqref{eq:res1} with $A^o_\alpha$ 
given by Eq.~\eqref{Ao} and $A^e_\alpha$ equal to
\begin{align}
\label{eq:2.E}
	A_\alpha^e &= 9\frac{\sqrt{5+ 2^{5-\alpha}/3}-3}{2(4-3\alpha)\log 2}.
\end{align}

The above results straightforwardly generalize to more involved tripartitions and can be used to access universal features 
of the negativity spectrum in the RSP. 
Let us recapitulate what one has to do in the most general case: 
(i) choose the partition, 
(ii) compute the average in-out singlets number solving the set of equations~\eqref{eq:1.B}, 
(iii) find the values of $p$ and $q$ using Eq.~\eqref{eq:2.X.1},
(iv) if $p,q\neq0$, then the negativity moments are just given by Eqs.~\eqref{Ee} and~\eqref{Eo}. 
The results will be valid only in the scaling regime of all length-scales  of the same order and much larger than $1$.

\section{Numerical tests for adjacent intervals}
\label{sec:NUM}
In this section we numerically test  the predictions reported above and in particular we compare our numerical simulations with the analytic formulas~(\ref{eq:Z1}-\ref{eq:Z4})
for the negativity moments of adjacent intervals.
We focus on the XX chain for which we can exploit known free fermion techniques 
to easily access the integer moments ${M^{T_2}_\alpha}$ with ab-initio simulations~\cite{Fagotti2010B,Coser2014,Eisler2014B,Coser2015}.
This kind of computations does not rely on the random singlet phase structure and thus represents a robust non-trivial check of our findings. 
We also implement numerically the SDRG providing another numerical benchmark which allows to explore much larger system sizes and  
easily access also the two analytic continuations of the moments to non-integer values.
All these simulations are extremely important in view of the several (reasonable) assumptions we made in writing down the renewal equation \eqref{renew}:
only the very good agreement between the predictions from its solution and the numerics represents a definitive confirmation for the correctness
of these assumptions, at least for asymptotically large RG time.

\subsection{Free fermions and negativity spectrum}
We review the mapping between the XX chain and free fermions on the line. 
Within the free fermions formalism, we can express the reduced density matrix and its partial transpose in the spin variables as a sum of Gaussian operators with known 
Majorana correlation matrices. 
The negativity moments are computed through a product rule for Gaussian matrices\cite{bb-69,Fagotti2010B,Coser2015}.

The Hamiltonian of the random XX chain is Eq.~\eqref{eq:2.0} at ${\Delta=0}$
\begin{equation}
\label{eq:APP1}
	H = \sum_{i=1}^{L-1} J_i(\sigma_i^x \sigma_{i+1}^x+ \sigma_i^y \sigma_{i+1}^y).
\end{equation}
Here we consider a chain of length $L$. The Jordan-Wigner transformation 
\begin{equation}
	c_i = \left(\prod_{m=1}^{i-1} \sigma^z_m\right) \frac{\sigma_i^x-i\sigma^y_i}{2},
\end{equation}
maps the Hamiltonian~\eqref{eq:APP1} in the free-fermion one
\begin{equation}
	\label{eq:APP2}
	H = \frac{1}{2}\sum_{i=1}^{L-1} J_i (c_i^\dagger c_{i+1} + c_{i+1}^\dagger c_i) \equiv \sum_{i,j=1}^{L-1} c_i^\dagger h_{i,j} c_j.
\end{equation}
The $c_i$ are fermion annihilation operator, satisfying the canonical anti-commutation relations ${\{c_m,c_n^\dagger\}=\delta_{mn}}$. 
At the free fermion point  the many-body eigenfunctions of $H$ can be expressed in terms of the single-particle ones $\{\phi_q(i) \}$;
the same is true for the many-body spectrum.  
Indeed, the eigenstates of the Hamiltonian~\eqref{eq:APP2} are obtained by applying an arbitrary number of single-particle creation operators 
\begin{equation}
	\eta_q^\dagger  = \sum_i \phi_q(i) c_i^\dagger,
\end{equation}
to the (fermonic) vacuum ${| 0 \rangle}$.
The ground state of \eqref{eq:APP1} corresponds to half-filling in fermionic language, that is the $N=L/2$ lowest energy levels are occupied
\begin{equation}
	|\textup{GS}\rangle = \eta_{q_N}^\dagger \cdots \eta_{q_1}^\dagger |0\rangle.
\end{equation}
The correlation matrix takes the form
\begin{equation} \label{cm}
	C_{ij} \equiv  \langle c_i^\dagger c_j\rangle = \sum_{q} \phi_q^*(i) \phi_q(j),
\end{equation}
where the sum is over the occupied single-particle excitations in the ground state. The reduced correlation matrix to a given subsystem $A$ with $\ell$ sites, ${C_A}$, is a ${\ell \times \ell}$ matrix whose elements are defined by the restriction of Eq.~\eqref{cm} to  ${i, j \in A}$.
For later convenience we also introduce the Majorana fermions
\begin{equation}
\label{eq:4.Q}
a_{2m -1} = c_m^{\dagger} + c_m, \quad 
a_{2m} = i (c_m^{\dagger} - c_m),
\end{equation}
and the corresponding ${2\ell \times 2\ell}$ Majorana correlation matrix $\Gamma_{A}$ with matrix elements
\begin{equation}
(\Gamma_{A})_{nm} \equiv \langle a_m a_n \rangle - \delta_{mn}.
\end{equation}
It is clear that there is a direct relation between the entries of $\Gamma_A$ and those of $C_A$.

Crucially, reduced density matrices associated with a single interval are Gaussian operators~\cite{Peschel2009}
\begin{equation}
\label{eq:4.Q.1}
\rho_A = \frac{1}{Z_\Omega} \exp \left( \frac{1}{4} \sum_{n, m} a_n \Omega_{nm} a_m \right),
\end{equation}
($Z_\Omega$ being a normalization) 
and the matrix $\Omega$ may be written in terms of $\Gamma_{A}$ as
\begin{equation}
\label{eq:4.Q.1B}
\Gamma_A \equiv \tanh \left( \frac{\Omega}{2} \right).
\end{equation}

However,  when the subsystem $A$ consists of more than one interval, the  reduced density matrix is not gaussian \cite{ip-09,atc-09}, 
and so also its partial transpose \cite{ez-15,Coser2015}.
Still, in both cases, the corresponding operator is the sum of gaussian terms. 
For instance, in the case of two disjoint intervals ${A=A_1\cup A_2}$, $\rho_A$ is the sum of two Gaussian operators $\rho_{\Gamma_i}$ associated 
by Eqs.~\eqref{eq:4.Q.1} and~\eqref{eq:4.Q.1B} to distinct covariance matrices $\Gamma_i$. 

With free fermion techniques is not straightforward to calculate the eigenvalues of the sum of Gaussian operators 
(see anyhow Ref. \onlinecite{zrc-19} for a brute force approach).
Instead, the traces of arbitrary integer powers of sums of (even non-commuting) gaussian operators can be calculated with, by now, 
standard methods \cite{bb-69,Fagotti2010B}. These methods heave been exploited already many times also for the calculation of negativity in 
spin chains \cite{ez-15,Coser2015,Nobili2015,ctc-16}.
Since the associated machinery is quite involved, here we just summarize the results and refer to the literature for further details \cite{ez-15,Coser2015}.
Let us denote by $ \rho_{\Gamma}$ the gaussian operator associated to the covariance matrix $\Gamma$.
Given $\Gamma$ and $\Gamma'$, we define the following product rule
\begin{equation}
\label{eq:prodQ}
\rho_{\Gamma} \rho_{\Gamma'}= {\rm Tr} \left[ \rho_{\Gamma} \rho_{\Gamma'}   \right]  \; \rho_{\Gamma \times \Gamma'},
\end{equation}
where  \cite{Fagotti2010B}
\begin{equation}
\Gamma \times \Gamma' \equiv 1- (1- \Gamma')\frac{ 1 }{1 + \Gamma \Gamma'} (1- \Gamma),
\end{equation}
relating the covariance matrices of two gaussian operators to the one associated to their product.
The trace on the right hand side of \eqref{eq:prodQ} is\cite{bb-69,Fagotti2010B}
\begin{equation}
 \{ \Gamma, \Gamma' \} \equiv {\rm Tr} \left( \rho_{\Gamma} \rho_{\Gamma'} \right) = \prod_{\mu \in \sigma(\Gamma \Gamma')/2} \frac{1+ \mu}{2},
\end{equation}
with the product being over half of the spectrum ${\sigma(\Gamma \Gamma')}$, which is doubly-degenerate.
Moreover, by associativity, one can extend this relation to more than two gaussian operators
\begin{equation}
\prod_{i=1}^n \rho_{\Gamma_{\alpha_i}} = \{  \Gamma_{\alpha_1} , \cdots , \Gamma_{\alpha_n}  \} \rho_{\Gamma_1 \times \cdots  \times \Gamma_{\alpha_n}},
\end{equation}
where
\begin{align} 
\nonumber
 \{ \Gamma_{\alpha_1}, \Gamma_{\alpha_2},\cdots, \Gamma_{\alpha_n} \} &\equiv {\rm Tr} \left( \rho_{\Gamma_{\alpha_1}}\rho_{\Gamma_{\alpha_2}}\cdots \rho_{\Gamma_{\alpha_n}} \right)\\\label{iteration}
   = \{ \Gamma_{\alpha_1} , &\Gamma_{\alpha_2} \} \{ \Gamma_{\alpha_1} \times \Gamma_{\alpha_2}, \cdots, \Gamma_{\alpha_n}\}.
\end{align}
The above equation can be used iteratively to evaluate traces of arbitrary products of gaussian operators.

In our case we need to identify the gaussian operators whose sum gives the partially transposed density matrix.
Let us specialize to the system studied in Sec.~\ref{subsec:adja} with two adjacent intervals $A_1$ and $A_2$, when there are major simplifications
compared to the case of disjoint intervals\cite{ez-15,Coser2015}.

Denoting with $\Gamma_{AA}$ the correlation matrix within ${A=A_1\cup A_2}$, 
we further define the four building blocks
\begin{align}
	\Gamma_1 &= \Gamma_{AA},\quad \Gamma_2 = P\Gamma_1 P,\\
	\tilde{\Gamma}_k &= \tilde{P} \Gamma_k\tilde{P} , \quad k=1,2, \label{blocksPT}
\end{align}
where
\begin{equation}
P=\begin{pmatrix}
 \mathbf{1}_{\ell} & 0  \\
 0 &  - \mathbf{1}_{\ell} \\
\end{pmatrix},
\quad {\rm and}\qquad
\tilde{P} =\begin{pmatrix}
 \mathbf{1}_{\ell} & 0  \\
 0 &  i\mathbf{1}_{\ell} \\
\end{pmatrix}.
\end{equation}
Here ${\mathbf{1}_{\ell}}$ is an identity matrix of dimension ${\ell\times\ell}$. 
The integer negativity moments may be written in terms of these building blocks~\cite{Coser2015}.
For convenience, we report here the cases of ${\alpha=2,3,4}$ which we use in the following
\begin{align}
 \label{eq:M2}
M^{T_2}_2 =& \{\tilde{\Gamma}_1,\tilde{\Gamma}_2\},\\
M^{T_2}_3 =& -\frac{1}{2}\{\tilde{\Gamma}_1,\tilde{\Gamma}_1,\tilde{\Gamma}_1\} + \frac{3}{2}\{\tilde{\Gamma}_1,\tilde{\Gamma}_1,\tilde{\Gamma}_2\},\\
M^{T_2}_4 =& -\frac{1}{2}\{\tilde{\Gamma}_1,\tilde{\Gamma}_1,\tilde{\Gamma}_1,\tilde{\Gamma}_1\} \nonumber\\
 \label{eq:M4}
	&+\frac{1}{2}\{\tilde{\Gamma}_1,\tilde{\Gamma}_2, \tilde{\Gamma}_1,\tilde{\Gamma}_2\}+\{\tilde{\Gamma}_1,\tilde{\Gamma}_1,\tilde{\Gamma}_2,\tilde{\Gamma}_2\}.
\end{align}
The above equations are used to numerically compute the disorder average of the negativity moments. 
The recipe is the following: (i) we choose a disorder realization of the free fermion single-particle Hamiltonian~\eqref{eq:APP2} with ${J_i\sim P(J)}$, 
(ii) we derive the correlation matrix for Majorana fermions $\Gamma_{AA}$, 
(iii) we construct $\Gamma_i$ and $\tilde \Gamma_i$ from the latter,
(iv) we compute the moments~(\ref{eq:M2}-\ref{eq:M4}),
and finally (v) iterating the process for many disorder realizations, we calculate the average. 

\begin{figure}[t]
\includegraphics[width=\columnwidth]{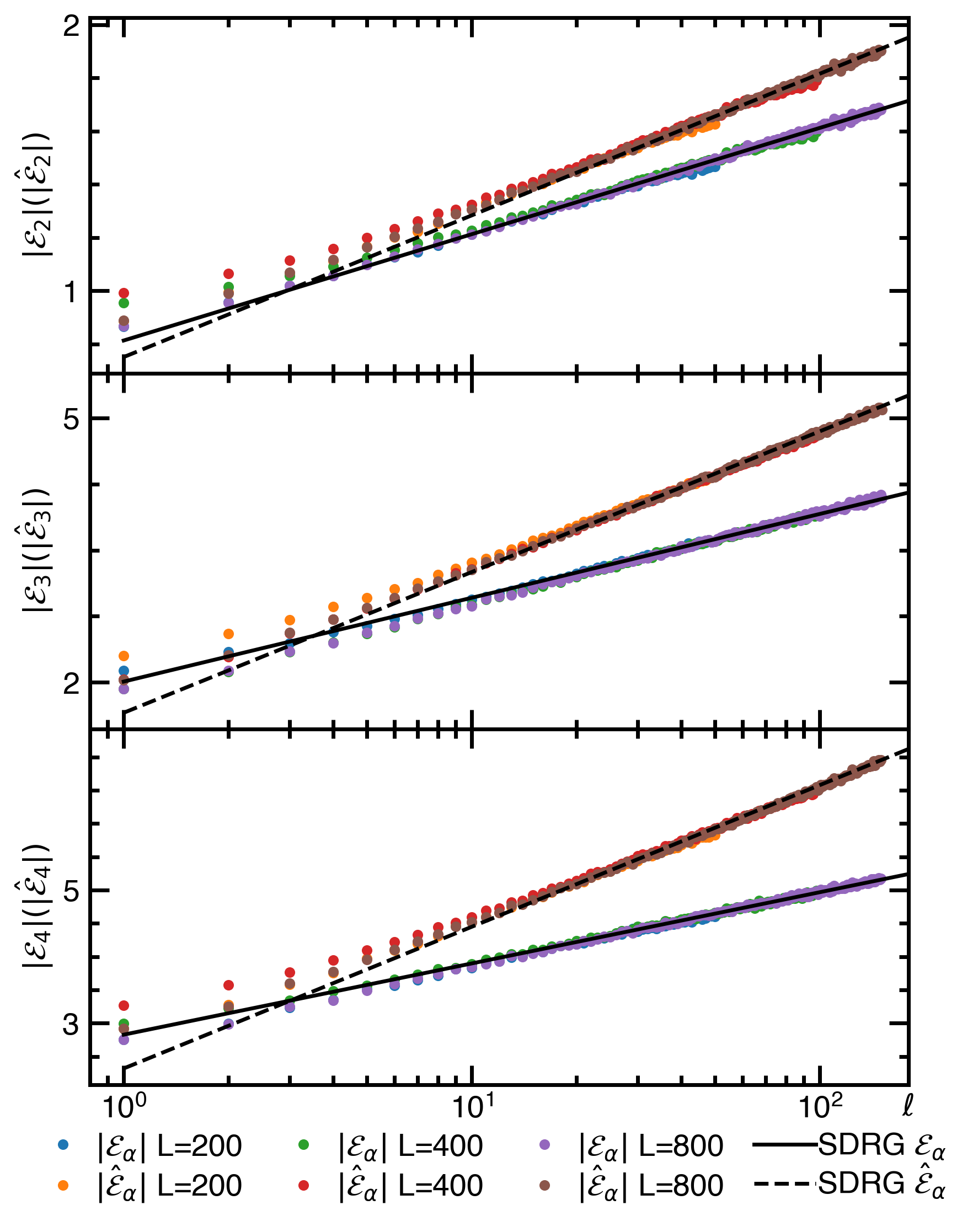}
\caption{Negativity moments ${\cal E}_\alpha$ and $\hat{\cal E}_\alpha$ for the random XX chain:  results of the \emph{ab initio} computations for two 
adjacent intervals of equal length $\ell$. 
We report the absolute values, since for the considered values of $\alpha$, they are all negative quantities (i.e. the moments $M^{T_2}_\alpha$ are smaller than $1$).
The symbols correspond to the numerical data while the continuous lines are the analytic SDRG predictions Eqs.~(\ref{eq:Z1}-\ref{eq:Z4}), 
with a best fit for the unknown non-universal additive constants. 
The agreement between the simulation and SDRG predictions is excellent already for moderate values of $\ell$. 
}
\label{Fig.2} 
\end{figure}

\subsection{Numerical results for adjacent intervals}
\label{subsec:nsdrg}

We are finally ready to test numerically the predictions reported in Sec.~\ref{subsec:adja}, as we do in the following. 
Throughout this section we consider a uniform coupling distribution ${P(J)=1}$ with ${J\in[0,1]}$, although, as stressed in Sec.~\ref{sec:RSP}, the results are 
distribution independent because of the universality of the RSP. 
In order to perform the numerical calculations, we must consider a finite chain of length $L$ and, for simplicity, we choose to work with open boundary conditions. 
In order to reduce the finite size and boundary effects we take the two adjacent intervals placed at the center of the chain.
We also limit our attention to the case of two intervals of equal length $\ell$, because all the universal factors may be extracted from this partition.

First we consider the \emph{ab-initio} method for XX chain, reviewed in the previous subsection. 
We focus on $\alpha=2,3,4$. 
We consider different system sizes $L=200,400,800$ and we vary the intervals length $\ell$ between ${\ell\in \{1,\dots,L/4\}}$.
We consider $10^5$ disorder realizations and we compute the disorder averages~\eqref{eq:2.A.4} and~\eqref{eq:2.A.4B}.
The obtained numerical data are reported in Fig. \ref{Fig.2}. 
It is evident that all ${\cal E}_\alpha$ grow logarithmically with $\ell$ as predicted.  
The logarithmic growth is compared with the analytic predictions in Eqs. (\ref{eq:Z1}-\ref{eq:Z4}). 
The agreement between the numerical data and SDRG is perfect already for moderate values of $\ell$.
In the plots the non-universal additive constants (not specified in Eqs. (\ref{eq:Z1}-\ref{eq:Z4})) have been fitted.   

\begin{figure}[t]
\includegraphics[width=\columnwidth]{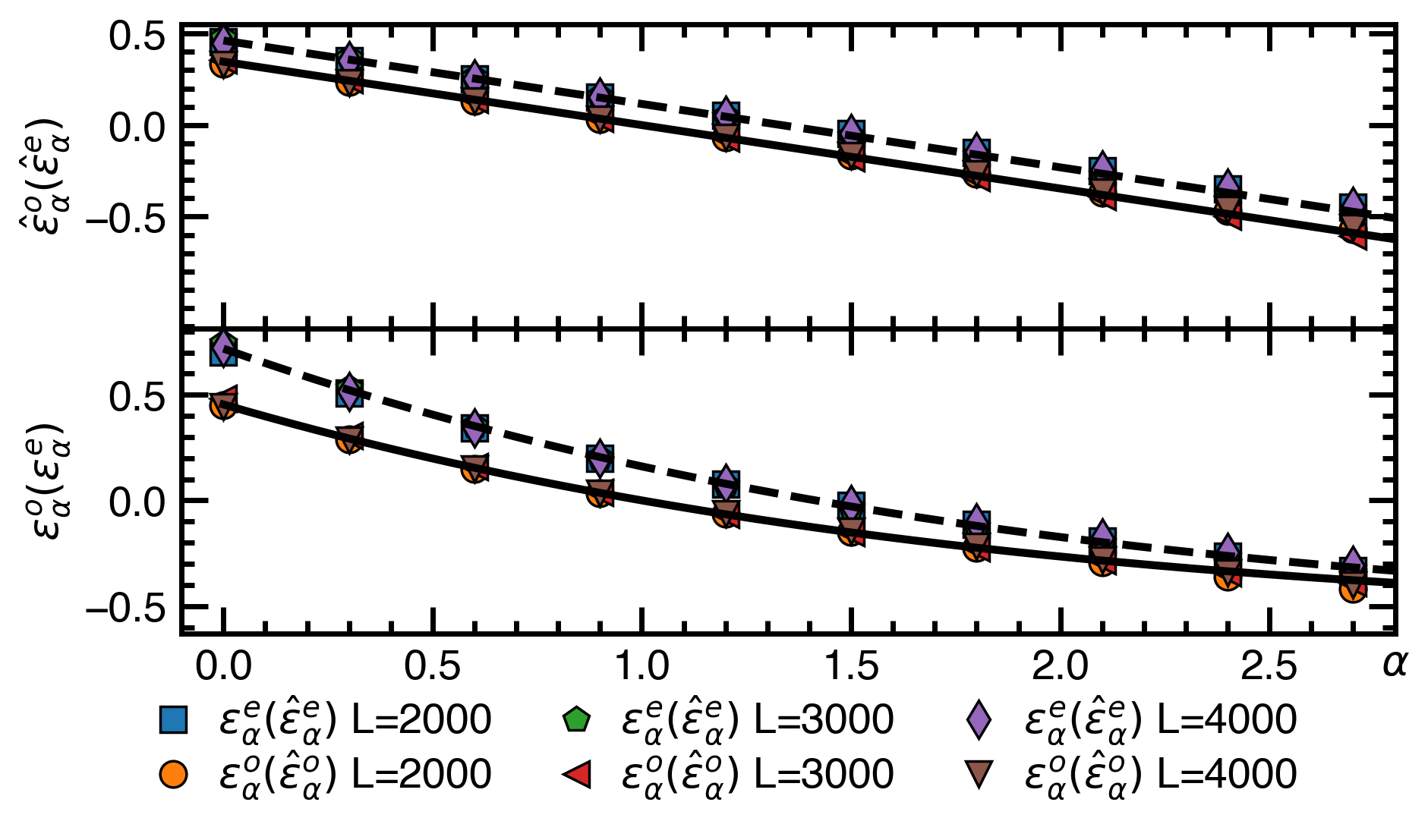}
\caption{\label{Fig.3} SDRG simulations for the negativity moments.
Here we plot, as function of $\alpha$, the prefactor  of the logarithm $\varepsilon^{e/o}_\alpha$ (cf.~\eqref{eps}), as obtained by a fit of the numerical data. 	
The symbols are the corresponding numerical data while the continuous lines are the analytic predictions in Eqs. (\ref{eq:Z1}-\ref{eq:Z4}). 
The plots show an extremely good agreement between numerical data and analytic predictions. 
}
\end{figure}

A byproduct of these ab-initio numerical simulations is an {\it indirect} test of the SDRG scaling for the logarithmic negativity 
obtained in Ref.~\onlinecite{Ruggiero2016B}. 
In fact, the latter is not efficiently accessed through free fermion techniques, because, as already stressed many times, the partially transposed reduced density matrix is not 
a non-gaussian operator. 
Therefore, via replica trick, the computation of $\hat {\cal E}_{2\alpha}$ provides an indirect check for the scaling of the logarithmic negativity as well.
This complement the numerical results obtained by SDRG and density matrix renormalization group in Ref. \onlinecite{Ruggiero2016B}.

We now implement numerically the SDRG for finite spin chains, defined by the Ma-Dasgupta rule~\eqref{eq:2.3}, which works as follows:
(i) pick up a random disorder realization with a list random couplings ${J_i\sim P(J)}$;
(ii)  iterate the Ma-Dasgupta rule (i.e. choose the strongest bond, build a singlet between them, remove the two sites, renormalize the coupling according to~\eqref{eq:2.3})
until all spins are paired up in singlets; at each step keep track of the location of the singlet and of the removed spins; 
(iii) count the in-out singlets formed between the partitions of interest; 
(iv) evaluate the negativity moments for the single realization using their form in terms of the number of singlets in Eq.~\eqref{eq:2.A.3}; 
(v) perform the average over all realizations. 
We vary the parameter $\alpha$, total length $L$, and the subsystem size $\ell$.  
Here the disorder average is taken over $10^6$ realizations. 
Since we are using Eq.~\eqref{eq:2.A.3} as operative definition of the negativity moments in the random singlet phase, 
we have direct access to the analytic continuations of all four families moments ${\cal E}^{e/o}_\alpha$ and $\hat {\cal E}^{e/o}_\alpha$
to even, odd and arbitrary non-integer values of $\alpha$.
From these numerical averages, we extract the prefactor of the logarithm for all the negativity moments for several values of $\alpha$
fitting the averages with
\begin{eqnarray}
\label{eps}
\hat{\cal E}_\alpha^{e/o}&=& \hat\varepsilon_\alpha^{e/o}\log\ell+ \hat k^{e/o}_\alpha,\\
{\cal E}_\alpha^{e/o}&=& \varepsilon_\alpha^{e/o}\log\ell+ k^{e/o}_\alpha . \nonumber
\end{eqnarray}
We restrict the fits to the windows of $\ell$ for which a logarithmic scaling is observed before finite size corrections kick in. 
The results for these four universal prefactors as function of $\alpha$ are reported in Fig. \ref{Fig.3}. 
The agreement between the analytic predictions  in Eqs.~(\ref{eq:Z1}-\ref{eq:Z4}) and the simulations is extremely good for the four moments and for all considered 
values of $\alpha$ (although some small finite size corrections are evident for the larger considered $\alpha$).
These SDRG results provide a test not only for the integer negativity moments, but also for their analytical continuations (\ref{eq:Z1}-\ref{eq:Z4}).

\section{Conclusion}
In this paper we exploited the Ma-Dasgupta decimation rule to write down a renewal equation for the probability distribution of in-out singlets in a tripartition of an infinite disordered spin chain in the random singlet phase.
This procedure assumes that the most relevant renormalization effect is a single decimation occurring at a specific bond.
The distribution resulting from the solution of the renewal equation provides analytic results for the negativity moments in the RSP and for their analytic continuations.
We focused on the case of adjacent intervals and the results have been numerically tested by means of \emph{ab-initio} simulations and numerical strong disorder renormalization group techniques, finding perfect agreement. 

Our analysis naturally rises a few questions deserving further investigations. 
The first one is that the true negativity spectrum (i.e. the full distribution of eigenvalues of $\rho_A^{T_2}$) has not yet been derived. 
Indeed, this is still and open issue also for the entanglement spectrum \cite{Fagotti2011} for which the calculation should be much simpler. 
A second natural question is to wonder whether it is possible to calculate the negativity moments for other infinite randomness fixed points that have been described in the literature. 

Finally, the dynamical evolution of the entanglement in random spin systems has been also subject to intensive investigation, 
especially in relation to many-body localized phases~\cite{Serbyn13,Vosk2014,Altman15,Nandkishore15,Parameswaran2017,Abanin18,Pekker2014,Zhao2016}.
A crucial aspect so far, even from the experimental side \cite{exp-lukin,exp-mbl},  has been to establish a quantitative understanding of the growth of the entanglement entropy.
It would be interesting to generalize some of these results to the negativity and negativity spectrum. 

\textbf{Acknowledgement} 
All authors acknowledge support from ERC: PC and PR  under Consolidator grant number 771536 (NEMO) and XT under Starting grant number 758329 (AGEnTh). 

\appendix

\section{Fermionic negativity moments}
\label{appendix}

The RSP describes also fermionic systems with random hoppings. 
However it has been shown was shown that the entanglement in the fermion variables is better captured by a \emph{fermionic negativity}, introduced in Ref.~\onlinecite{ssr-17} and related to a partial time reversal operation.  
The associated spectrum has been studied for disorder-free fermions~\cite{Hassan2019A}.
Therefore it is interesting to understand this spectrum in random systems and particularly in the RSP.
In this appendix, we recall the definition of the two possible density matrices for fermionic negativity and determine them 
in the random singlet phase.

For interacting fermions, the hamiltonian is obtained from the XXZ hamiltonian \eqref{eq:2.0}, via a Jordan-Wigner trasformation
\begin{equation}
H = \sum_j J_j \left( c_j^{\dagger} c_{j+1}+ c_{j+1}^{\dagger} c_j - \Delta n_j + \Delta n_j n_{j+1} \right),
\end{equation}
where $c_j, c_j^{\dagger}$ are spinless fermionic operators and ${n_j = c_j^{\dagger} c_j}$ the occupation number of the $j$-th site of the chain.
The non-locality of such transformation points at a modification of the SDRG prescription to take into
account the fermionic nature of the particles. 
It was shown~\cite{sierra-fermions} that this can be implemented through a simple modification of the RG prescription as
\begin{equation} \label{DMrule-fermions}
\tilde{J} = - \frac{J_L J_R}{\Omega}.
\end{equation}
Eq.~\eqref{DMrule-fermions} implies that the hoppings can now be either positive or negative. 
When they are positive, a singlet-type bond is established between two sites, of the form ${| \psi_- \rangle \propto |01 \rangle - | 10 \rangle}$, written in the occupation number basis of the fermions.
If the hopping is negative, the corresponding triplet-type anti-bond is established ${| \psi_+ \rangle \propto |01 \rangle + | 10 \rangle}$.
A crucial point is that the two types of bonds share many properties, such as entanglement.
In particular, the spectrum of the associated density matrices is the same, ${\sigma (\rho_+)= \sigma (\rho_-)}$, where ${\rho_{\pm} = | \psi_{\pm} \rangle \langle \psi_{\pm}|}$.
The same is true for the corresponding (standard) partial transpose,  ${\sigma (\rho^{T_2}_+)= \sigma (\rho^{T_2}_-)}$.
Therefore, for our purpose, the ground state in the RSP can be written as
\begin{equation}
|GS \rangle = \prod_{i} | \psi_{-} \rangle_i .
\end{equation}

In the occupation number basis, the fermionic partial trasponse differs from the standard partial transpose just by a phase $e^{i \pi \phi}$, with
\begin{multline}
 \phi (  \{ n_j \} , \{ \bar{n}_j \} )=  \frac{\tau_1 (\tau_1+1)}{2} + \frac{\bar{\tau}_1 (\bar{\tau}_1+1)}{2} + 
\tau_2 \bar{\tau}_2 \\
+ \tau_1 \tau_2 + \bar{\tau}_1 \bar{\tau}_2 + (\tau_1 + \tau_2) (\bar{\tau}_1 + \bar{\tau}_2).
\end{multline}
Here ${\tau_s =\sum_{i \in A_s} n_i}$, ${\bar{\tau}= \sum_{i \in A_s} \bar{n}_i}$ and refer to the ket ${| \{ n_j \} \rangle}$ and bra ${\langle \{ \bar{n}_j\} |}$ state, respectively. See Ref.~\onlinecite{ssr-17} for details.
In particular, applying the definition to our building block $\rho_-$, with the two subsystems consisting of a single site each, leads to
\begin{align} 
\rho_{-}^{R_2} =
&\frac{1}{2}
\begin{pmatrix}
0 & 0   & 0 & - i\\
0 & 1   & 0 & 0\\
0 & 0  &  1 & 0\\
- i & 0  &  0& 0
\end{pmatrix},
\label{R2}
\end{align}
whose spectrum is given by ${\{ i/2 ,-i/2, 1/2, 1/2 \}}$.

We can now apply Eq. \eqref{R2} to the reduced density matrix of the RSP, after tracing $B$, i.e.,
\begin{equation}
	\rho_A = \bigotimes_{m=1}^{n_{A:A}}\rho_\textup{2s}\bigotimes_{n=1}^{n_{A:B}}\rho_\textup{s}, 
	\nonumber
\end{equation}
where here ${\rho_{\textup{2s}} = \rho_-}$ and ${\rho_{\textup{s}} = \textrm{tr}_s \rho_-}$ (with the trace being on one of the two sites). We obtain
\begin{equation}
\label{rhoR2tot}
	\rho_A^{R_2}= 
\left\{	\prod_{k=1,2} \bigotimes_{p=1}^{n_{A_k:A_k}} \rho_\textup{2s} \right\}
	\bigotimes_{q=1}^{n_{A_1:A_2}}\rho^{R_2}_\textup{2s}
\left\{	\prod_{k=1,2} \bigotimes_{r=1}^{n_{A_k:B}}\rho_\textup{s} \right\}.
\end{equation}
Here we have used the fact that ${\rho_\textup{s}^{R_2}= \rho_\textup{s}}$ for a single site, and that ${\rho_\textup{2s}^{R_2}= \rho_\textup{2s}}$ when both the ends of a bond are in the same subsystem $A_i$ $(i=1,2)$.

There are 4 different non-zero eigenvalues
\begin{equation} \label{lambda-untwisted}
\lambda_k = 2^{-n_{A:B} - n_{A_1:A_2} } e^{i k \pi/2}, \; k=0, \pm1, 2.
\end{equation}

These come with degeneracies $d_k$ given by
\begin{align} \label{deg-untwisted}
& d_{\pm1}  = 2^{n_{A:B} +  n_{A_1: A_2} -2} \left( 2^{ n_{A_1: A_2}} \right), \nonumber \\
& d_{0} = 2^{n_{A:B} + n_{A_1 :A_2} -2 }  \left( 2^{n_{A_1: A_2}} +2 \right) ,  \\
& d_{2} = 2^{n_{A:B} + n_{A_1 :A_2} -2 }  \left( 2^{n_{A_1: A_2}} -2 \right). \nonumber
\end{align}
From \eqref{lambda-untwisted} and \eqref{deg-untwisted} we notice that the moments, ${M_{\alpha}^{R_2} \equiv \textrm{tr}\left( \rho_A^{R_2} \right)^{\alpha}}$ have three different analytic continuations when restricting to $\alpha$ integer
\begin{equation} \label{untwistedM}
M_{\alpha}^{R_2} = 
 2^{(n_{A:B} + n_{A_1 : A_2}) (1 - \alpha)} 
\begin{cases}
1  \qquad \qquad & \alpha = 2p+1,\\ 
 2^{n_{A_1 :A_2}} \qquad &\alpha = 4p,\\
0 \qquad \qquad & \alpha = 4p +2, \\
\end{cases}
\end{equation}
with $p$ integer. This more complicated periodicity has already been observed in the translational invariant setting~\cite{Hassan2019A}.
As a check, from the odd sequence in \eqref{untwistedM} we recover the proper normalization 
${\textrm{tr} \rho^{R_2}_A = \lim_{\alpha \to 1 } M_{\alpha}^{R_2} =1}$.

This was dubbed \emph{untwisted} negativity spectrum in Ref.~\onlinecite{Hassan2019A}, with the important difference with respect to the standard negativity spectrum of spin and bosonic models, of being complex.
On the other hand, also for fermions one can introduce a hermitian partial transpose, more suitable to define another fermionic negativity due to its real spectrum. 
This is done by considering the composite operator ${\rho_{\times} \equiv ( \rho^{R_2}_A )^{\dagger} \rho^{R_2}_A}$ and by noting that ${\rho_{\times} = ( \rho^{\tilde{R}_2}_A)^{2}}$, where we introduced the \emph{twisted} partial transpose ${\rho^{\tilde{R}_2}_A \equiv \rho_A^{R_2} (-1)^{F_2}}$ of Ref.~\onlinecite{Hassan2019A}. Here ${(-1)^{F_2}}$ is the fermion number parity in $A_2$, since ${F_2 = \sum_{j \in A_2} n_j }$. 

For the RSP, the twisted partial transposed reads
\begin{equation}
\label{rhoR2tottil}
	\rho_A^{\tilde{R}_2}= 
\left\{\prod_{k=1,2} \bigotimes_{p=1}^{n_{A_k:A_k}} \rho_\textup{2s} \right\}
	\bigotimes_{q=1}^{n_{A_1:A_2}}\rho^{\tilde{R}_2}_\textup{2s}
\left\{	\prod_{k=1,2} \bigotimes_{r=1}^{n_{A_k:B}}\rho_\textup{s} \right\}.
\end{equation}
It has two non-zero eigenvalues
\begin{equation}
\tilde{\lambda}_{\pm}=\pm 2^{- n_{A:B} - n_{A_1 :A_2}},
\end{equation}
with equal degeneracy
\begin{equation}
\tilde{d}_{\pm}= 2^{ n_{A:B} +2 n_{A_1 :A_2}-1}.
\end{equation}
Therefore, the associated moments, $M^{\tilde{R}_2}_{\alpha} \equiv \textrm{tr} \left( \rho^{\tilde{R}_2}_A \right)^{\alpha} $, are given by
\begin{equation} \label{twistedM}
M^{\tilde{R}_2}_{\alpha} = 
\begin{cases}
0 \hspace{4.2cm} \alpha \, \textrm{odd} \\
2^{(n_{A:B} + n_{A_1 : A_2})(1-\alpha) }   2^{n_{A_1 :A_2}} \quad \alpha \, \textrm{even} .
\end{cases}
\end{equation}
The negativity is obtained from Eq.~\eqref{twistedM} via replica limit as 
\begin{equation}
\mathcal{E} = \lim_{\alpha \to 1/2} \overline{\log M_{2 \alpha}^{\tilde{R}_2}} = \overline{n_{A_1 :A_2}} \log 2.
\end{equation}
Actually, in this case, Eq.~\eqref{untwistedM} also implies that
\begin{equation}
\mathcal{E} = \lim_{\alpha  \to 1/4}  \overline{\log M_{4 \alpha}^{{R}_2}}  = \overline{n_{A_1 :A_2}} \log 2.
\end{equation}
Note that, as already shown numerically in Ref.~\onlinecite{ssr-17}, this means that in the case of fermions we recover the result obtained for the equivalent spin system in Ref.~\onlinecite{Ruggiero2016B}.

From Eqs.~\eqref{untwistedM} and \eqref{twistedM}, it is clear that the fermionic negativity spectrum is different from the corresponding one in the spin variables. In fact, there are integer values of $\alpha$ for which they are trivial, i.e. are exactly vanishing. Nevertheless, the non-trivial moments have the same functional forms of the moments~\eqref{eq:2.A.3}. As such, the same techniques employed in Section~\ref{sec:ENS} may be used to obtain SDRG results for the  disordered-average moments associated to twisted and
untwisted density matrices. For example, within the same assumptions of Sec.~\ref{sec:ENS}, the non-trivial untwisted moments reads
\begin{equation} \label{untwistedAve}
\log\overline{M_{\alpha}^{R_2}} = 
\begin{cases}\displaystyle
 \frac{\sqrt{5+2^{3-\alpha}} - 3}{2}\mu + \dots \qquad \qquad & \alpha = 2p+1,\\ 
\displaystyle \frac{\sqrt{5+(1+q)2^{3-\alpha}} - 3}{2}\mu + \dots \qquad &\alpha = 4p,\\
\end{cases}
\end{equation}
while the non-trivial twisted ones are
\begin{equation}
	\log\overline{M_{\alpha}^{\tilde{R}_2}}=\frac{\sqrt{5+(1+q)2^{3-\alpha}} - 3}{2}\mu, \qquad \alpha \ \text{even}.
\end{equation}


\end{document}